\documentclass[10pt]{article}

\usepackage{latexsym}

\font\sf=cmss10 at 10pt 

\font\bb=msbm10 at 10pt

\font\frak=eufm10 at 10pt
\font\fraks=eufm10 at 8 pt
\font\cal=cmsy10 at 9pt
\font\cals=cmsy10 at 7pt

\def\0#1{\mbox{\rm#1}}
\def\1#1{\mbox{\bb#1}}
\def\2#1{\mbox{\bf#1}}
\def\3#1{\mbox{\cal#1}}
\def\4#1{\mbox{\cals#1}}
\def\5#1{\mbox{\sf#1}} 
\def\6#1{\mbox{\frak#1}}
\def\7#1{\mbox{\fraks#1}}
\def\8#1{{\tilde #1}}
\def\9#1{{\breve #1}}

\def\BEn{\begin{enumerate}}
\def\EEn{\end{enumerate}}
\def\BEq{\begin{equation}}
\def\EEq{\end{equation}}
\def\BEqA{\begin{eqnarray}}
\def\EEqA{\end{eqnarray}}

\def\de{\delta}
\def\De{\Delta}

\def\la{\lambda}

\def\si{\sigma}

\def\tav{\hbox{
\kern-1.0pt
\rule[0pt]{1.3pt}{.8pt}{\kern-3.6pt}
\rule[0pt]{.4pt}{6pt}{\kern-3.1pt}
\rule[4.5pt]{3.0pt}{.8pt}{\kern-3.3pt}
\rule[0pt]{.4pt}{5pt}{\kern-1pt}
}}

\def\lfl{{\lfloor\kern-5pt\lfloor}}
\def\rfl{{\rfloor\kern-5pt\rfloor}}

\def\pd{\partial}

\def\from{\kern-2pt\leftarrow\kern-2pt}

\def\Bar{\kern5pt{\rule[-2.5pt]{.6pt}{9.5pt}}\kern5pt}

\def\II{|\kern-1pt |}

\begin{document}

\title{ {\bf Unimodular relativity and
cosmological constant}\\
(To appear in the {\it Journal of Mathematical Physics}) }

\author{David R.
Finkelstein, Andrei A. Galiautdinov, and James E. Baugh\\
\small{\it School of Physics, Georgia Institute of
Technology, Atlanta, Georgia 30332-0430}}

\maketitle

\abstract{Unimodular relativity is a theory of gravity
and space-time with a fixed
absolute space-time volume
element, the {\em modulus}\/, which we suppose is
proportional to the number of microscopic modules in that
volume element. In general relativity an arbitrary fixed
measure can be imposed as a gauge condition, while in
unimodular relativity it is determined by the events in the
volume.
Since this seems to break general covariance,
some have suggested that it permits a non-zero
covariant divergence of the material stress-energy tensor
and a variable cosmological ``constant.''
In Lagrangian unimodular relativity, however,
even with higher-derivatives of the gravitational field
in the dynamics,
the usual covariant continuity holds
and the cosmological constant
is still a constant of integration of the gravitational
field equations.}



\section{Introduction to unimodular relativity}

Unimodular relativity is an alternative theory of gravity
considered by Einstein in 1919 \cite{EINSTEIN}
without a Lagrangian and put into Lagrangian form by Anderson and
Finkelstein \cite{AF}.  The space-time of unimodular
relativity is a measure manifold, a manifold provided by
nature with a fixed absolute physical measure field
$\mu(x)$ to be found by direct measurement,
subject to no dynamical development.
The sole structural variable is a conformal metric
tensor $f_{\mu\nu}$, subject to dynamical equations.
The measure of a space-time region may be regarded
as indirectly counting the modules
of which it is composed,
in the way that the volume of a lake indirectly
counts its water molecules.
Both space-time measure and liquid measure
indicate a modular structure
below the
limit of resolution
of the present measuring instruments.

The
conformal metric field
$f_{\mu\nu}(x)$ is a symmetric relative tensor of
weight $1/2$, signature $1-3$,
and determinant $-1$ in all coordinate systems,
with 9 independent
components, operationally defined by the system of light paths, whose
tangent vectors
$dx^{\mu}$ obey
$f_{\mu\nu}dx^{\mu}dx^{\nu}=0$.

The unimodular space-time structure also defines a metric
tensor
\BEq
\label{eq:GMN}
g_{\mu\nu}=\sqrt{\mu}\,f_{\mu\nu}(x),
\EEq
but  the determinant
\BEq
\label{eq:DET}
-g:=\det
g_{\mu\nu}=-\mu^2
\EEq
is not a dynamical variable.
The conformal metric $f$
is the sole gravitational variable
of unimodular relativity.

We assume that the metric tensor field $g_{\mu\nu}$
found by measuring the proper times
$d\tau^2=  g_{\mu\nu}dx^{\mu}dx^{\nu}$
for a sufficiently fine network of
intervals $dx^{\mu}$,
also determines the measure field
$\mu$ by the usual relation (\ref{eq:DET}).

Once the measure $\mu$ has been experimentally determined
it establishes
a class of admissible metrics obeying (\ref{eq:DET}).
Metrics violating (\ref{eq:DET}) are unphysical
according to unimodular relativity.

The variable of general relativity is a compound
of a light-cone field $f$ and a measure field $\mu$, and the group of
general relativity is a non-simple group of diffeomorphisms,
with an invariant subgroup of unimodular coordinate
transformations, those with Jacobian $\det(\partial
x'/\partial x)=1$. Unimodular relativity has a simple group
and a simple variable.

Originally
we proposed unimodular relativity because there is
indeed an experimental atomic standard of length
near each point of space-time, not built into
general relativity \cite{AF} .
This suggests that
the macroscopic structure of space-time
is a smoothed description of an underlying
atomic space-time microstructure,
which seems necessary for other reasons.

Since the actual value of the cosmological constant is so
finely tuned, it is natural to attempt to
derive its value from physical principles.
A theory in which it is
variable  would be a useful starting point
for any such attempt.
Recently the difference in symmetry between
unimodular and general relativity led some to hope that
they might differ on the constancy of this parameter
  \cite{TIWARI}.

On the other hand, many authors
have already argued that
the difference is only a gauge condition,
which has no physical consequences \cite{ALL}. However, some
authors do not share this point of view. In particular, van der
Bij et al \cite{BIJ} stated very clearly the physical
difference between the usual formulation of gravity and the
unimodular theory. Also, an interesting and somewhat alternative
approach is presented in \cite{SORKIN}.

We examine the gauge-condition argument more carefully here.
In its usual form it omits several relevant features
special to this problem.
Usually gauge conditions are applied to Lagrangians
that are already physically well-defined in their absence;
the unimodularity condition is not
a gauge condition of this kind.

One should also take into account the possibility of higher-order
derivatives in the gravitational equations,
of the kind that might arise from renormalization
in some hypothetical quantum field theory of gravity.

We show here that any gravitational theory of classical unimodular
relativity  with a Lagrangian density
that is invariant under the unimodular coordinate group
is
equivalent in its experimental predictions to
a theory of classical general
relativity.
Higher-order corrections do not disturb this equivalence.

\section{The metric tensor of unimodular relativity}

In deriving the field equations from a variational
principle (on which our approach is based),
the measure
$\mu$ is not varied but is treated as if it were a
fixed external field.
This disturbs general covariance.
The law of nature may take a simpler form
in {\em unimodular coordinates}\/, where $\mu(x)\equiv 1$.
Unimodular coordinate systems
are related by unimodular transformations.

Let $R$ be the Riemann scalar computed from the metric tensor
$g_{\mu\nu}$ of (\ref{eq:GMN}).
Let ${L}_M$ be the Lagrangian density of the matter field in
the presence of $g_{\mu\nu}=\sqrt{\mu}f_{\mu\nu}$.
Then
\BEq
\label{eq:URACTION}
S =  \int  d^4 x \, ({\Gamma\over 2} R + L_M)
\EEq
is a possible action functional for unimodular relativity
in a unimodular coordinate system.
The constant $\Gamma=1/4\pi G$
is the inverse rationalized gravitational constant,
the  reciprocal square of the rationalized Planck length, in
units with $\hbar = c =1$.

In unimodular  relativity, there is initially
no way to vary all 10
components of
$g_{\mu\nu}$ independently.
The action is in principle defined only for $g=\mu^2$.
Only derivatives with respect to the
9-dimensional conformal metric field $f$ are defined.

A cosmological constant term $\Lambda \sqrt{g}$ in the action
function would be an ineffectual additive constant
since $\sqrt{g}=\mu$ is not varied.

This action can be transformed to any other coordinate system
under the general diffeomorphism group,
but is not generally invariant
in functional form,
since the fixed measure $\mu$ sets an absolute scale
at each event.

The derivative with respect to the conformal metric $f$ requires special
care. Since infinitesimal variations $\de f$ are subject to the unimodular
condition (\ref{eq:DET}), they obey
\BEq
\label{eq:DF}
f^{\mu\nu}\de f_{\mu\nu}=0 \,.
\EEq
If
$W:\3F\to \3W, \quad f\mapsto W(f)$ is a functional from the
function manifold
$\3F$ of conformal metrics on a region $\3R$
to some value-manifold $\3W$, we define the functional derivative
$W_f=\delta W/\delta f$
as the linear operator
\BEq
W_f: d\3F\to \3W
\EEq
such that for any $\delta f \in d\3F$
vanishing on the boundary $\pd \3R$
the tangent space to $\3F$ at $f$,
\BEq
\delta W= { W_f}\cdot \delta f\,=
\int d^4x \,{\de W\over \de f_{\mu\nu}(x)} \de
f_{\mu\nu}(x) \,.
\EEq
Then the dynamical equation that follows from the action principle
for any space-time region $\3R$ is
\BEq
\de S(\3R)= \int_{\pd
\4R}{\kern-5pt}d\si_{\mu}\;\pi^{\mu}\cdot\de f \,.
\EEq
with a boundary term that is linear in
    $\de f$ on the boundary $\pd \3R$
and vanishes for variations that vanish on the boundary.
The tensor field $\pi^{\mu}$ canonically conjugate to $f(x)$ is defined by
these relations.

\section{Field equations and the cosmological constant}

It is often inconvenient to work with a field variable
subject to non-linear conditions like the conformal metric.
One may
reformulate unimodular relativity
with an unconstrained variable $g_{\mu\nu}(x)$
and take the unimodular condition
(\ref{eq:DET})
into account through Lagrange's method of undetermined multipliers.

To do this, however,
we must give values to the Lagrangian density
for metric tensors that have $g\ne \mu^2$
(\cite{AF}, \cite{ALL}, \cite{NG}), and hence
are unphysical.
One way to do this is to replace $\mu$ by $\sqrt{g}$ and
$f_{\mu\nu}$ by $g_{\mu\nu}/g^{1/4}$
at all their ``appearances'' in the action $S$\/,
so that $S$ is defined
in a 10-dimensional neighborhood $\3F'\supset\3F$ in a
smooth way consistent with the values on $\3F$.
Then in addition one adds
a Lagrangian-multiplier term
expressing the unimodular condition.
We call the resulting extension of $S$ to $\3F'$ the {\em
extended action function}
$S'$.

But this prescription is ambiguous,
since it depends on ``appearances,''
on how $S$ is written, on matters of notation.
The extended action $S'$ is arbitrary up to
a correction term $\Delta_M S$
depending on the matter variables and the metric
$g_{\mu\nu}$,
subject only to the condition that
$\Delta_M{L}$ and its derivative with
respect to $\lambda$
vanish in the
unimodular sector  (\ref{eq:DET}):
\BEq
\label{eq:AA}
S'= S+\De_M S + \int  d^4x \, \sqrt{g} \, \la
(x)[{\mu\over
\sqrt{ g(x)}}-1]
\EEq
$\Delta_M S=\int d^4 x \, \sqrt{g} \, \Delta_M L
$ is the {\em unimodular ambiguity} in the action.
No physical results may depend on the choice of
$\Delta_M L$,
and so no physical experiments can determine $\Delta_M L$.

In mechanical theories sometimes we impose a constraint
and thereby reduce an already
well-defined system to a system of lower
dimensionality.
For example, we reduce a free particle
to a spherical pendulum by constraining it to a sphere.
Then there is a well-defined unconstrained Lagrangian,
found by removing the constraint, and
the ambiguity $\Delta_M L $
does not arise.

But according to unimodular relativity
we have no way of actually removing
the unimodular condition (\ref{eq:DET}).
In this sense it is not a constraint,
so we call it a condition.
While the proper time $d\tau$ of a coordinate interval
$dx^{\mu}$ at $x$ depends on the gravitational field
at $x$, each coordinate  cell $d^4 x$
at $x$ comes with its own intrinsic measure
$\mu(x)d^4 x$, independent of gravity.
The unimodular ambiguity $\Delta_M L$
acknowledges that
as a matter of principle we cannot know
how the system
would evolve absent the unimodular condition.

The admissible infinitesimal variations $df$
within the neghborhood $\3F'$
are those that
  obey
(\ref{eq:DF})
for all $x\in \3R$
That is, they all belong to the null space
of the inverse conformal metric tensor $f^{\mu\nu}$.

The action principle states that for any physical field
$f=(f_{\mu\nu}(x))$,
and for any variations $\de f=(\delta f_{\mu\nu}(x))$
about
$f$,
\BEq
S_f\cdot \delta f:= \int_{\4R} d^4 x \, S_f(x)\de f(x)
=\int_{\pd \4R}  S^{\mu\nu}\de f_{\mu\nu}.
\EEq

That is, any vector $\de f(x)$ in the null space of
$f^{-1}(x)$ is in the null space of $S_f(x)$.
It follows that $S_f(x)$ is in the ray of $f^{-1}(x)$:
\BEq
{\delta S\over \de f_{\mu\nu}(x) }= \lambda(x) f^{\mu\nu}(x).
\EEq
The multiplier $\lambda(x)$ is then fixed by the unimodular condition.

This implies that the augmented action
(\ref{eq:AA})
is stationary up to
boundary terms when we vary
$\la(x)$ and the 10 components
$g_{\mu\nu}(x)$ independently.

The unimodular condition makes
the  stress tensor ambiguous as well as the dynamical equations.
In unimodular  relativity the general relativistic
concept of the stress tensor
\BEq
\label{eq:STRESS}
{T^{\mu \nu}}
= \frac{1}{\sqrt{g}}\frac{\delta(\sqrt{g}{L}_M)}{\delta
g_{\mu\nu}}
\EEq
has no principled meaning at first, since it involves
breaking the unimodular condition, nor has
the statement of covariant continuity,
$T^{\mu \nu}{}_{; \nu}=0$.

We may suppose that $\Delta_M L$
has the form
\BEq
\Delta_M
{L}=\left[\frac{\mu(x)}{\sqrt{g(x)}}
- 1\right] l_M,
\EEq
where $l_M$ is any function of the matter variables and
$g_{\mu\nu}$. We write
${L}'_M:={L}_M+\Delta_M L$ for the sum.

  From any general-relativity action principle $S$  we
obtain in this way an ambiguous unimodular-relativity
action principle
\BEq
\label{eq:AMBACTION}
S' =  \int  \sqrt{ g(x)} \;d^4x  \left\{
{\Gamma\over 2}
R(x)  + {\Gamma\over 2} \lambda(x)
\left[\frac{\mu(x)}{\sqrt{g(x)}}-1\right]
+ {L}_M +\Delta_{M} {L}\right\}.
\EEq
The second term in $S'$ expresses the unimodular
condition
and breaks general covariance.
The fourth term expresses the unimodular ambiguity.
We vary the 10
$g_{\mu\nu}(x)$ and the Lagrange multiplier $\lambda(x)$
independently.
We have written the $\la$ term in a form that makes $\lambda$
a scalar field under the general group.

Variation of
$\lambda $ in $S'$ recovers the condition
\BEq
\label{eq:VARLAMBDA}
\sqrt{g(x)} = \mu(x)
\EEq
The unimodular ambiguity $\Delta_M L $ does
not affect this result,
since it vanishes when the unimodular condition holds.

Variation of $g_{\mu \nu}$ gives the equation of motion
\BEqA
\delta S' = \int \sqrt{g} \; d^4x \; \left\{
\Gamma \left(\frac{1}{2} g^{\mu \nu} R
- R^{\mu \nu} \right)
+ \frac{1}{2}\Gamma \lambda g^{\mu \nu} \right\}
\delta g_{\mu \nu} \cr
+\int \sqrt{g} \; d^4x \; \left\{ \frac{1}{\sqrt{g}}
\frac{\delta}{\delta g_{\mu
\nu}}(\sqrt{g} {L}_M+\sqrt{g}\Delta_{M} L) \right\}
\delta g_{\mu \nu}=0,
\EEqA
or
\BEq
\label{eq:VARGMUNU}
      R^{\mu \nu} - \frac{1}{2} g^{\mu
\nu} R  -
\frac{1}{2} \lambda  g^{\mu \nu} = 8 \pi G T'^{\mu \nu},
\EEq
where $G:=1/ \Gamma$ is a rationalized gravitational coupling strength
and
\BEq
\label{eq:SET}
T'^{\mu \nu} := \frac{1}{\sqrt{g}}
\frac{\delta}{\delta g_{\mu \nu}} (\sqrt{g} {L}_M+\sqrt{g}
\Delta_M L)\/.
\EEq

Tracing (\ref{eq:VARGMUNU}) gives  the Lagrange multiplier:
\BEq
\lambda = - \frac{1}{2}(8 \pi GT' + R).
\EEq
The field equation then simplifies to
\BEq
\label{eq:FIELDEQS}
R^{\mu \nu} - \frac{1}{2} R g^{\mu \nu} = 8\pi G T'^{\mu \nu}
-
\frac{1}{4} g^{\mu \nu} (8\pi GT' + R),
\EEq
or equivalently
\BEq
\label{eq:FIELDEQSTRACELESS}
R_{\mu \nu} - \frac{1}{4} R g_{\mu \nu} = 8\pi G T'{}^{\rm
T}_{\mu \nu},
\EEq
where
\BEq
\label{eq:SESS}
T'{}^{\rm T}_{\mu \nu} := T'_{\mu \nu} - \frac{1}{4}
g_{\mu\nu} T'
\EEq
is the traceless (part of the) stress tensor, or the {\em
sess tensor} (note that the tr has been removed).
The covariant divergence of (\ref{eq:FIELDEQS}) is
\BEq
\label{eq:COVDIVFORT}
8\pi G {T'^{\mu \nu}}_{; \nu} = \frac{1}{4} g^{\mu
\nu}(8\pi G T' + R)_{, \nu} \equiv -\frac{1}{2} g^{\mu \nu}
\lambda,
\EEq
which was suggested as a ``modified covariant divergence
law'' by Tiwari \cite{TIWARI}.

In general relativity, general invariance of the action
function implies that  the covariant divergence
${T'^{\mu \nu}}_{; \nu}$ vanishes in virtue of the field
equations for matter \cite{LANDAU}, \cite{MTW}.
Then it follows that
\BEq
-\frac{1}{2}(8\pi GT' + R) = \lambda = {\rm constant}.
\EEq
If the stress tensor of unimodular
relativity were covariantly continuous
too, then the undetermined multiplier $\lambda$
could be identified with a cosmological constant $\Lambda$,
though now a constant of the motion determined by the initial data,
rather than
a fixed absolute constant as supposed in general
relativity.

In unimodular relativity however, $T'^{\mu\nu}$ is
not covariantly continuous because the action $S'$ is not
generally covariant, which seems to justify Tiwari's
suggestion. $T'^{\mu\nu}$ is ambiguous by an additive term
  \BEq
  \Delta T^{\mu \nu} = - \frac{1}{2} g^{\mu \nu} l_M
  \EEq
  and its trace is correspondingly ambiguous by
  \BEq
  \Delta T = -2\, l_{\rm M}.
  \EEq
However substitution of these expressions into
(\ref{eq:FIELDEQSTRACELESS}) immediately leads to
\BEq
\label{eq:FIELDEQSTRACELESSNEW}
R_{\mu \nu} - \frac{1}{4} R g_{\mu \nu} = 8\pi G T{}^{\rm
T}_{\mu \nu},
\EEq
with
\BEq
\label{eq:SESSNEW}
T{}^{\rm T}_{\mu \nu} := T_{\mu \nu} - \frac{1}{4}
g_{\mu\nu} T,
\EEq
where $T_{\mu \nu}$ is the usual covariantly continuous
stress-energy tensor (\ref{eq:STRESS}) of the matter field.
The gravitational field equations {\it do not} depend on the
ambiguity in the matter field Lagrangian. The cosmological
constant again arises as a constant of integration.

Einstein's law of gravity is a second-order differential equation
for the metric field. In quantum field theories higher-order
derivatives arise from renormalization.
One might hope that in higher-order theories,
the unimodular relativity differs in content from general relativity,
and allows the comsological "constant" to vary.
In such theories the cosmological constant
may be defined as $\la=S(h)/\mu$ where $S(h)$ is the value of the
gravitational action density for the case of the Minkowski metric.

It easily follows from the generalized
(contracted) Bianchi identities of the higher order theory
(see, for example, \cite{QUERELLA}, \cite{WALD}) that even in
higher-order unimodular relativity,
where the gravitational part of the action is an arbitrary
generally invariant functional of the curvature scalar, the
cosmological constant also arises as the constant of
integration.

\section{Conclusions}

We have shown that in Lagrangian unimodular relativity
the usual covariant continuity equation holds for
the source stress tensor,
and the cosmological constant
is a constant of integration of the gravitational
field equations.
Higher-derivatives of the gravitational field
may appear in the Lagrangian without
disturbing these conclusions.
The essential point is that
the stress tensor have no covariant divergence.
This follows from either unimodular or general covariance.

There are several reasons not to
be quite certain that these classical conclusions will still hold
in the quantum domain.

The fact remains that unimodular
relativity  forces us to allow many values as possibilities
for the cosmologcal constant,
while general relativity fixes on one value.
In quantum theories,
possibilities affect actualities.
In a quantum theory of sufficient scope,
these many possibilities for the cosmological constant
of unimodular relativity might influence
the actual experimental situation \cite{NG}.

Furthermore, quantum field theories
often lack symmetries and conservation laws
present in the classical
Lagrangian from which they stem,
due to divergences inherent in the limiting process
used to define the quantum theory.
This results in quantum anomalies, for example.
Similar effects may permit the cosmological constant ---
the vacuum rest-energy-density ---
to vary in some quantum version of
unumodular relativity.

Perhaps the most basic weakness in the deduction
is the postulate of strong locality.
This is implicit in general covariance
and is required to deduce the covariant
continuity of stress.
If there is a fundamental quantum time,
a limit to locality,
as some suggest,
then the cosmological constant can vary.

\section*{Acknowledgments}

This work was aided by discussions with N. Dragon, R. Sorkin,
C. Teitelboim and E. Witten. It was partially supported by the
M. and H. Ferst Foundation.


\begin{thebibliography}{References}

\bibitem{EINSTEIN}
A. Einstein, Sitzungsber. d. Preuss. Akad. d. Wissench., Pt.
1, 433 (1919). English translation in H.A. Lorentz, A.
Einstein et al., {\it The Principle of Relativity}, Dover,
New York (1952)

\bibitem{AF}
J. Anderson, D. Finkelstein, {\it Amer. J. Phys.},
{\bf 39/8}, 901 (1971)

\bibitem{TIWARI}
S.C. Tiwari, J. Math. Phys. {\bf 34} (6), 2465 (1993)



\bibitem{ALL}
F. Wilczek, Phys. Rep. {\bf 104}, 111 (1984);
A. Zee, in {\it High Energy Physics}, proceedings of the
20th Annual Orbis Scientiae, Coral Gables, 1983, edited by
B. Kursunoglu, S.C. Mintz, and A. Perlmutter, Plenum, New
York (1985); W. Buchm{\"u}ller, N. Dragon, Phys. Lett. {\bf
B207}, 292 (1988); W.G. Unruh, Phys. Rev. {\bf D40}, 1048
(1988); M. Henneaux, and C. Teitelboim, Phys. Lett. {\bf B222},
195 (1989); L. Bombelli et al., Phys. Rev. {\bf D44}, 2589
(1990)

\bibitem{BIJ}
J.J. van der Bij et al., Physica {\bf A116}, 307 (1982)

\bibitem{SORKIN}
R.D. Sorkin,
Int. J. Theor. Phys. {\bf 33}, 523 (1994);
``Problems with Causality in the Sum-over-histories
Framework for Quantum Mechanics'',
    in A. Ashtekar and J. Stachel (eds.),
   {\it Conceptual Problems of Quantum Gravity}, proceedings
of the conference held Osgood Hill, Mass., May 1988,
217, Boston, Birkh\"auser, (1991);
Int. J. Theor. Phys. {\bf 36}: 2759 (1997), also
    LANL gr-qc/9706002;
A. Daughton, J. Louko and R.D. Sorkin,
Phys.Rev. {\bf D58}: 084008 (1998), also LANL
gr-qc/9805101

\bibitem{NG}
Y.J. Ng, and H. van Dam, J. Math. Phys. {\bf 32} (5),
1337 (1991), LANL hep-th/9911102

\bibitem{LANDAU}
L.D. Landau, E.M. Lifshitz {\it The Classical Theory of
Fields}, 4th revised english ed., Pergamon Press
(1975),  p. 270

\bibitem{MTW}
C.W. Misner, K.S. Thorne, and J.A. Wheeler {\it
Gravitation}, Freeman, San Francisco (1973)


\bibitem{QUERELLA}
L. Querella, {\it Variational Principles and Cosmological
Models in Higher-Order Gravity}, Doctoral dissertation,
Universite de Liege (1998)


\bibitem{WALD}
R.M. Wald, {\it General Relativity}, The University of
Chicago Press (1984), Appendix E.


\end{thebibliography}
\end{document}